\documentclass[twocolumn, twocolappendix]{aastex631}
\usepackage{amsmath}
\usepackage{enumitem}
\usepackage{newtxtext,newtxmath}
\usepackage{bm}
\usepackage{multirow}
\usepackage{xurl}
\usepackage{anyfontsize}
\usepackage{needspace}
\usepackage{latexml}
\iflatexml\else
\makeatletter
\let\orig@startsection\@startsection
\def\@startsection#1#2{\ifnum#2<4 \needspace{4\baselineskip}\fi\orig@startsection{#1}{#2}}
\makeatother
\fi
\hypersetup{breaklinks,colorlinks,citecolor=blue,urlcolor=blue,linkcolor=red}

\newcommand{\uatentry}[2]{\href{http://astrothesaurus.org/uat/#2}{#1 (#2)}}

\usepackage{soul}
\soulregister{\ref}{7}
\soulregister{\citep}{7}
\soulregister{\citet}{7}
\soulregister{\textit}{7}
\soulregister{\texttt}{7}

\makeatletter\renewcommand\@makefntext[1]{\noindent\@makefnmark#1}\makeatother
 
\graphicspath{{./}{figures/}}

\shorttitle{\it AM-1 stream in Euclid DR1}
\shortauthors{\it Chen et al.}

\begin{document}

\title{\vspace{-6mm}\large A stellar stream from the distant Galactic globular cluster AM-1 discovered in \textit{Euclid} DR1}

\author[0000-0002-5970-2563]{Yingtian Chen}
\affiliation{Center for Cosmology and AstroParticle Physics (CCAPP), The Ohio State University, 191 West Woodruff Ave, Columbus, OH 43210, USA}
\affiliation{Department of Astronomy, The Ohio State University, 140 West 18th Ave, Columbus, OH 43210, USA}
\correspondingauthor{Yingtian Chen}
\email{chen.15712@osu.edu}

\author[0000-0001-9852-9954]{Oleg Y. Gnedin}
\affiliation{Department of Astronomy, University of Michigan, 1085 South University, Ann Arbor, MI 48109, USA}

\author[0000-0003-0872-7098]{Adrian M. Price-Whelan}
\affiliation{Center for Computational Astrophysics, Flatiron Institute, 162 Fifth Ave, New York, NY 10010, USA}

\author[0000-0001-5082-9536]{Yuan-Sen Ting}
\affiliation{Department of Astronomy, The Ohio State University, 140 West 18th Ave, Columbus, OH 43210, USA}
\affiliation{Center for Cosmology and AstroParticle Physics (CCAPP), The Ohio State University, 191 West Woodruff Ave, Columbus, OH 43210, USA}
\affiliation{Max-Planck-Institut f\"ur Astronomie, K\"onigstuhl 17, D-69117 Heidelberg, Germany}

\begin{abstract}
The first data release (DR1) of the \textit{Euclid} mission became public in November 2026, providing deep imaging over $\sim 1900\ {\rm deg^2}$ of sky area. Its depth enables searches for faint stellar streams around distant globular clusters (GCs), most of which lack detected streams. In this work, we search for a stream around Arp--Madore~1 (AM-1), one of the most distant Galactic GCs at $D_\odot\approx 120$~kpc, and the only known Galactic GC in the \textit{Euclid} Deep Fields. We adapt the stream detection method \textit{StarStream} to \textit{Euclid} photometry, using Gaussian mixture models to model the stream in sky positions, colors, and magnitudes. We detect a stream at a formal significance of $4.1\sigma$, including 15 stars with a stream membership probability above 0.5. The stream has two short tails, $\sim 0.3$~deg long, aligned with the cluster proper motion. Most members lie near or below the main-sequence turnoff ($I_E \approx 24$). Based on the density of this stream, we infer a mass loss rate $\dot{M}=0.24^{+0.12}_{-0.09}\ M_\odot\, {\rm Myr^{-1}}$ averaged over $t_{\rm dur}\approx1.1$~Gyr, consistent with theoretical predictions based on $N$-body simulations.
\end{abstract}



\section{Introduction}

Stellar streams are elongated structures of stars tidally stripped from their progenitors, such as globular clusters (GCs) and dwarf galaxies (DGs) \citep{lynden-bell_ghostly_1995}. The stripped stars escape through the Lagrange points and spread roughly along the progenitor's orbit, forming leading and trailing tails \citep{kupper_more_2012, fardal_generation_2015}. The first GC streams were found in wide-field photometric surveys, such as the stream of Palomar~5 \citep{odenkirchen_detection_2001, rockosi_matched-filter_2002} and GD-1 \citep{grillmair_detection_2006}. With the \textit{Gaia} mission \citep{gaia_collaboration_gaia_2016}, the number of known streams in the Milky Way (MW) halo has rapidly grown to over 100 \citep{malhan_ghostly_2018,shih_via_2021,ibata_charting_2021, ibata_charting_2024,bonaca_stellar_2025,chen_starstream_2026}. Since streams trace the orbits and mass loss of their progenitors, they are powerful probes of the MW potential, dark matter substructure, and the dynamical evolution of GCs \citep{koposov_constraining_2010,bovy_shape_2016,erkal_number_2016,kupper_globular_2015,gieles_supra-massive_2021,chen_stellar_2025}.

However, among the more than 150 Galactic GCs \citep{hilker_galactic_2019}, fewer than 40 have been confidently associated with streams \citep{chen_starstream_2026}. This ``missing stream'' puzzle has two possible explanations. First, most GCs may not produce detectable streams. This is either because their mass loss rate is lower than expected, or because stellar-mass black holes or internal dynamics make their tails diffuse \citep{gieles_mass-loss_2023,weatherford_stellar_2024,roberts_stellar_2025}. Second, their streams may simply be too faint for current surveys \citep{pearson_forecasting_2024,holm-hansen_catalog_2026}. To resolve this puzzle, we need surveys that are both deep and wide.

The recent \textit{Euclid} mission \citep{euclid_collaboration_euclid_2025} provides one such survey. Its Wide Survey will cover $\sim14,000\ {\rm deg}^2$ \citep{euclid_collaboration_euclid_2022}, and its sharp imaging with the Visible Instrument (VIS) reliably separates faint stars from galaxies down to $I_E\approx25$. \textit{Euclid} also repeatedly observes the \textit{Euclid} Deep Fields (EDFs), which were first released at wide-survey depth in the Quick Data Release 1 \citep[Q1;][]{euclid_collaboration_euclid_2026}. Arp--Madore~1 \citep[AM-1;][]{madore_three_1979} is the only known Galactic GC in the EDFs. It is one of the most distant MW GCs at $D_\odot\approx 120$~kpc, with its main-sequence turnoff (MSTO) at $I_E \approx 24$, beyond the reach of \textit{Gaia} and any other prior surveys in its vicinity. Among them, the Dark Energy Survey (DES) is closest, with a median $10\sigma$ depth of $i \approx 23.8$ \citep{abbott_dark_2021}. No stream has been reported around it. \citet{weisz_dolphoteuclid_2026} recently studied resolved stars in AM-1 with \textit{Euclid} Q1 imaging. The deeper \textit{Euclid} Data Release 1 (DR1) of EDF-South now allows to search for its stream.

In this work, we search for a stream around AM-1 in \textit{Euclid} DR1 with a physics-based stream detection method, \textit{StarStream} \citep{chen_starstream_2025}. After adapting this method to \textit{Euclid} photometry, we detect a stream signal at $4.1\sigma$. This paper is organized as follows. We describe the data in \S\ref{sec:data} and the detection method in \S\ref{sec:method}. We present the detection results in \S\ref{sec:results}. We then provide a straightforward matched-filter view of this stream (\S\ref{sec:window}) and infer the mass loss rate of AM-1 (\S\ref{sec:mdot}). Finally, we summarize our findings in \S\ref{sec:summary}.

\section{Data}
\label{sec:data}

\subsection{AM-1 data}

Our data for the AM-1 cluster come from two sources. The right ascension (RA), declination (Dec), proper motion (PM; $\mu_\alpha^*$, $\mu_\delta$)\footnote{$\mu_\alpha^*\equiv\mu_\alpha\cos\delta$}, heliocentric distance ($D_\odot$), and line-of-sight velocity ($v_{\rm los}$) are taken from the fourth edition (March 2023) of the online Galactic GC database\footnote{\url{https://people.smp.uq.edu.au/HolgerBaumgardt/globular/}} maintained by Baumgardt et al. \citep[see][]{baumgardt_accurate_2021,vasiliev_gaia_2021,baumgardt_absolute_2020,baumgardt_evidence_2023,sollima_eye_2019}. The stellar population age, $\rm [Fe/H]$, dust extinction ($A_V$), and stellar mass ($M_\star$) are taken from \citet{weisz_dolphoteuclid_2026}, whose isochrone parameters come from fits to archival \textit{Hubble Space Telescope} (\textit{HST}) photometry. We list the above quantities in Table~\ref{tab:am1}. 

\begin{table}
    \centering
    \caption{Adopted and derived key properties of AM-1.}
    \label{tab:am1}
    \renewcommand{\arraystretch}{1.05}
    \begin{tabular}{lrl}
        \hline\hline
        Quantity & Value & Unit \\
        \hline
        \multicolumn{3}{l}{\textit{\href{https://people.smp.uq.edu.au/HolgerBaumgardt/globular/}{Galactic GC database}}} \\
        RA & $58.7596$ & deg \\
        Dec & $-49.6153$ & deg \\
        $\mu_{\alpha}^{*}$ & $0.384 \pm 0.066$ & mas\,yr$^{-1}$ \\
        $\mu_{\delta}$ & $-0.196 \pm 0.082$ & mas\,yr$^{-1}$ \\
        Heliocentric distance ($D_\odot$) & $118.9 \pm 3.4$ & kpc \\
        Line-of-sight velocity ($v_{\rm los}$) & $118.0 \pm 14.1$ & km\,s$^{-1}$ \\
        Mass function slope ($\alpha$) & $-1.07 \pm 0.34$ & \vspace{1mm} \\
        \hline
        \multicolumn{3}{l}{\textit{\citet{weisz_dolphoteuclid_2026}}} \\
        Age & $10.6 \pm 0.6$ & Gyr \\
        $[{\rm Fe/H}]$ & $-1.00 \pm 0.07$ & dex \\
        $A_V$ & $0.022$ & mag \\
        Stellar mass ($M_\star$) & $2.1 \times 10^{4}$ & $M_\odot$ \\
        Half-light radius ($r_{\rm h}$) & $12.5^{+0.8}_{-0.7}$ & pc \vspace{1mm} \\
        \hline
        \multicolumn{3}{l}{\textit{This work}} \\
        $\mu_{\alpha}^{*}$ & $0.42^{+0.06}_{-0.05}$ & mas\,yr$^{-1}$ \\
        $\mu_{\delta}$ & $-0.20^{+0.03}_{-0.04}$ & mas\,yr$^{-1}$ \\
        Stream release duration ($t_{\rm dur}$) & $1.1^{+0.7}_{-0.5}$ & Gyr \\
        Mass loss rate ($\dot{M}$) & $0.24^{+0.12}_{-0.09}$ & $M_\odot$\,Myr$^{-1}$ \vspace{1mm} \\
        \hline\hline
    \end{tabular}
\end{table}

The PM from the Galactic GC database has large uncertainties at such a long distance. The solar reflex motion alone is $(\mu_\alpha^*, \mu_\delta) \approx (0.29, -0.15)\ {\rm mas\,yr^{-1}}$ at the position of AM-1. After subtracting it, both $\mu_\alpha^*$ and $\mu_\delta$ are consistent with being at rest within 1--2$\sigma$. Therefore, we do not directly use the database PM for stream modeling. Instead, we set PM as free model parameters and sample them with Markov chain Monte Carlo (MCMC; \S\ref{sec:method_mcmc}). 

\begin{figure}
    \centering
    \includegraphics[width=\linewidth]{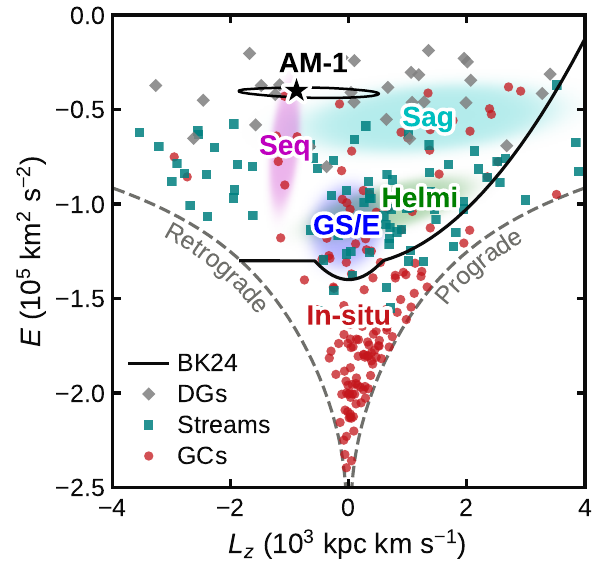}
    \vspace{-5mm}
    \caption{Atlas of the MW's satellite DGs (gray diamonds, adopted from \citealt{Pace2025}), streams (teal squares, adopted from \citealt{bonaca_stellar_2025}), and GCs (red circles, adopted from the Galactic GC database) in the IoM space: total orbital energy $E$ vs. the $z$-component of the angular momentum $L_z$. For comparison, we plot the in-situ/ex-situ boundary by \citet{belokurov_-situ_2024} as the solid curve. The AM-1 cluster is shown as the black star symbol, with $E$ and $L_z$ uncertainties represented by the $1\sigma$ credible region (black ellipse). We derive the credible region from the Galactic GC database's reported values and errors, with PM replaced by our new measurements (\S\ref{sec:results}). We show the approximate locations of major merger events as blurred ellipses, whose centroids and spreads are derived from the GC classifications by \citet[][GS/E and Sagittarius]{chen_galaxy_2024} and \citet[][Helmi and Sequoia]{massari_origin_2019}.}
    \label{fig:am1_iom}
\end{figure}

To better understand the dynamical origin of AM-1, we show in Fig.~\ref{fig:am1_iom} its location in the integrals-of-motion (IoM) space: total orbital energy $E$ vs. the $z$-component of the angular momentum $L_z$. We compute AM-1's location in the IoM space along with the $1\sigma$ credible region\footnote{Throughout this paper, the $1\sigma$ credible region of a 2D distribution refers to the region enclosing 39.3\% of the probability, the same fraction as within $1\sigma$ of a 2D Gaussian.} via Monte Carlo sampling. The sample is generated from a multivariate Gaussian distribution based on the uncertainties reported in the Galactic GC database, except that we use our own best-fit PM (\S\ref{sec:results}). 

In the IoM space, AM-1 is among the highest-energy GCs with orbits similar to satellite DGs. Its angular momentum is negative (we define $L_z>0$ as prograde), reflecting its polar, slightly retrograde orbit. However, the angular momentum has a large spread due to the uncertain PM. The $1\sigma$ credible region extends to positive $L_z$, meaning that AM-1's orbit may be prograde. The orbit has a moderate eccentricity $\approx0.6$, with an apocenter radius $\approx130$~kpc and a pericenter radius $\approx30$~kpc to the Galactic center. The cluster's current distance indicates that it is near its apocenter. 

For comparison, we also show in Fig.~\ref{fig:am1_iom} other Galactic GCs (from the same database), satellite DGs (adopted from \citealt{Pace2025}), and stellar streams \citep[adopted from][]{bonaca_stellar_2025}. The orbital energies for these objects are computed using the \texttt{MilkyWayPotential2022} model \citep{price-whelan_gala_2017}, the same as in our later stream modeling (\S\ref{sec:method}). This potential is very similar to the one used by \citet{belokurov_-situ_2024}. Therefore, we can directly compare the distributions of these objects with \citet{belokurov_-situ_2024}'s boundary for accreted/in-situ origins. To subdivide the region for accreted objects into individual merger events, we also show the approximate locations of four major mergers: \textit{Gaia}-Sausage/Enceladus (GS/E), Sagittarius, Helmi, and Sequoia. The centroids and spread of these events in the IoM space are derived from the Galactic GC classifications by \citet{chen_galaxy_2024} and \citet{massari_origin_2019}. We emphasize that the subdivision of the IoM space is only for guiding visual interpretation and should not be treated as a strict classification.

AM-1 is consistently classified as an accreted GC by \citet{chen_galaxy_2024}, \citet{massari_origin_2019}, and \citet{belokurov_-situ_2024}. However, it does not lie close to the centroid of any of the major merger events. AM-1's $1\sigma$ credible region marginally touches the outskirts of Sagittarius and Sequoia in the IoM space, indicating that AM-1 may be loosely connected to either of these events. But we cannot rule out the possibility that AM-1 was accreted through its own merger event.

We neglect the effect of the LMC. In a test with a moving LMC of $1.38\times10^{11}\ M_\odot$ \citep{erkal_total_2019} added to the MW potential, the axis of the simulated stream in the search region turns by only 1~deg.

\newcommand{\FigFootprint}{%
\begin{figure}
    \centering
    \includegraphics[width=\linewidth]{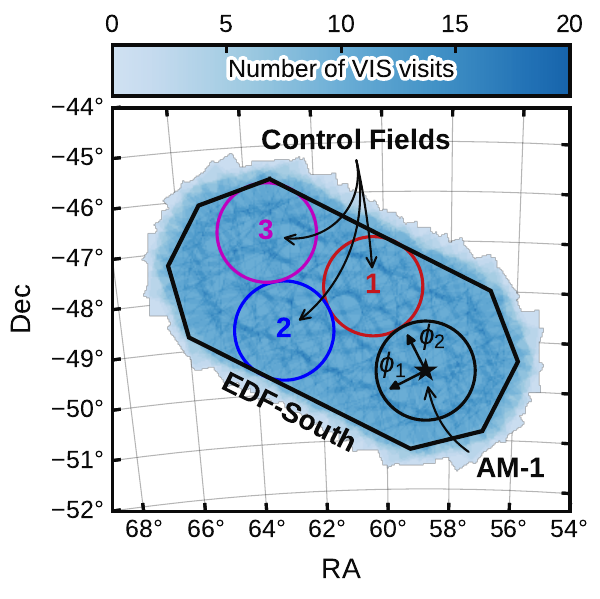}
    \vspace{-5mm}
    \caption{AM-1's location (star symbol) in the EDF-South footprint (solid octagon). The 1-deg search radius around AM-1 is shown as the solid circle. The three control fields of the same radius are shown as the colored circles. We color-code each pixel by the number of VIS visits in \textit{Euclid} DR1.}
    \label{fig:euclid_dr1_footprint_edfs}
\end{figure}}
\iflatexml\else\FigFootprint\fi

\subsection{\textit{Euclid} data}

In November 2026, the European Space Agency (ESA) publicly released \textit{Euclid} DR1, which includes $\sim1900\ {\rm deg^2}$ of sky area from observations taken between spring 2024 and spring 2025 across the \textit{Euclid} Wide and Deep Surveys. Our target, AM-1, is the only known Galactic GC in the EDFs. As shown in Fig.~\ref{fig:euclid_dr1_footprint_edfs}, EDF-South fully covers AM-1, including its 1-deg vicinity where its potential stream resides. In EDF-South, \textit{Euclid} has reached an average of $\sim10$ VIS visits in DR1. As a result, the $5\sigma$ depth for point sources reaches $I_E\approx28$~mag and $Y_E$, $J_E$, $H_E\approx26$~mag (all in the AB system).

We use the MER catalog of EDF-South and retrieve all sources within 1~deg of AM-1. We take $I_E$ from the VIS point spread function (PSF)-fitting flux and $Y_E$ from the Near-Infrared Spectrometer and Photometer (NISP) template-fitting flux. We keep sources detected in VIS, not flagged as spurious, with positive fluxes and uncertainties in both bands. We select point sources with $\mu_{\rm max} - m < -2.8$, where $\mu_{\rm max}$ is the peak surface brightness and $m$ is the magnitude (both in $I_E$) just above the stellar locus at $\approx -3.0$. We require $I_E<25$ to robustly remove faint, compact galaxies that dominate $I_E\gtrsim26$, while this cut is already 1~mag below AM-1's MSTO. We further require $|I_E - Y_E| < 5$ to remove extremely red/blue sources. For the later stream search, we also mask the cluster itself by removing sources within the tidal radius (311~pc, or 0.150~deg). We take the tidal radius to be the Jacobi radius at AM-1's present position and velocity, computed for a mass of $2.1\times10^4\,M_\odot$ in the \texttt{MilkyWayPotential2022} model. A total of 30,987 sources pass all cuts. The point-source cut alone removes $>95\%$ of all sources. Hereafter, we refer to these point sources as stars, although a few may be unresolved galaxies.

Our selection prioritizes completeness. The conservative $I_E<25$ and $|I_E - Y_E| < 5$ cuts ensure high and uniform stellar completeness up to the magnitude limit. Note that we mask out the cluster itself, so crowding does not affect completeness. This is by design to obtain an unbiased estimate of the cluster mass loss rate in \S\ref{sec:mdot}. However, several factors may still lower the completeness. The dominant one is the point-source cut, which may exclude $\sim10\%$ of real stars. This loss percentage does not correlate with magnitude. For sources matched to \textit{Gaia} Data Release 3 \citep[$I_E\lesssim21$;][]{gaia_collaboration_gaia_2023}, the completeness is 90--91\%, matching the loss percentage.

We also define three control fields of the same 1-deg radius in EDF-South, spaced by 2~deg so that they do not overlap with each other (see Fig.~\ref{fig:euclid_dr1_footprint_edfs}). We apply the identical selection to them. The locations of these fields are chosen arbitrarily since they are only used to perform null tests for stream detection.

\iflatexml\FigFootprint\fi

\newcommand{\FigCorner}{%
\begin{figure*}
    \centering
    \includegraphics[width=\linewidth]{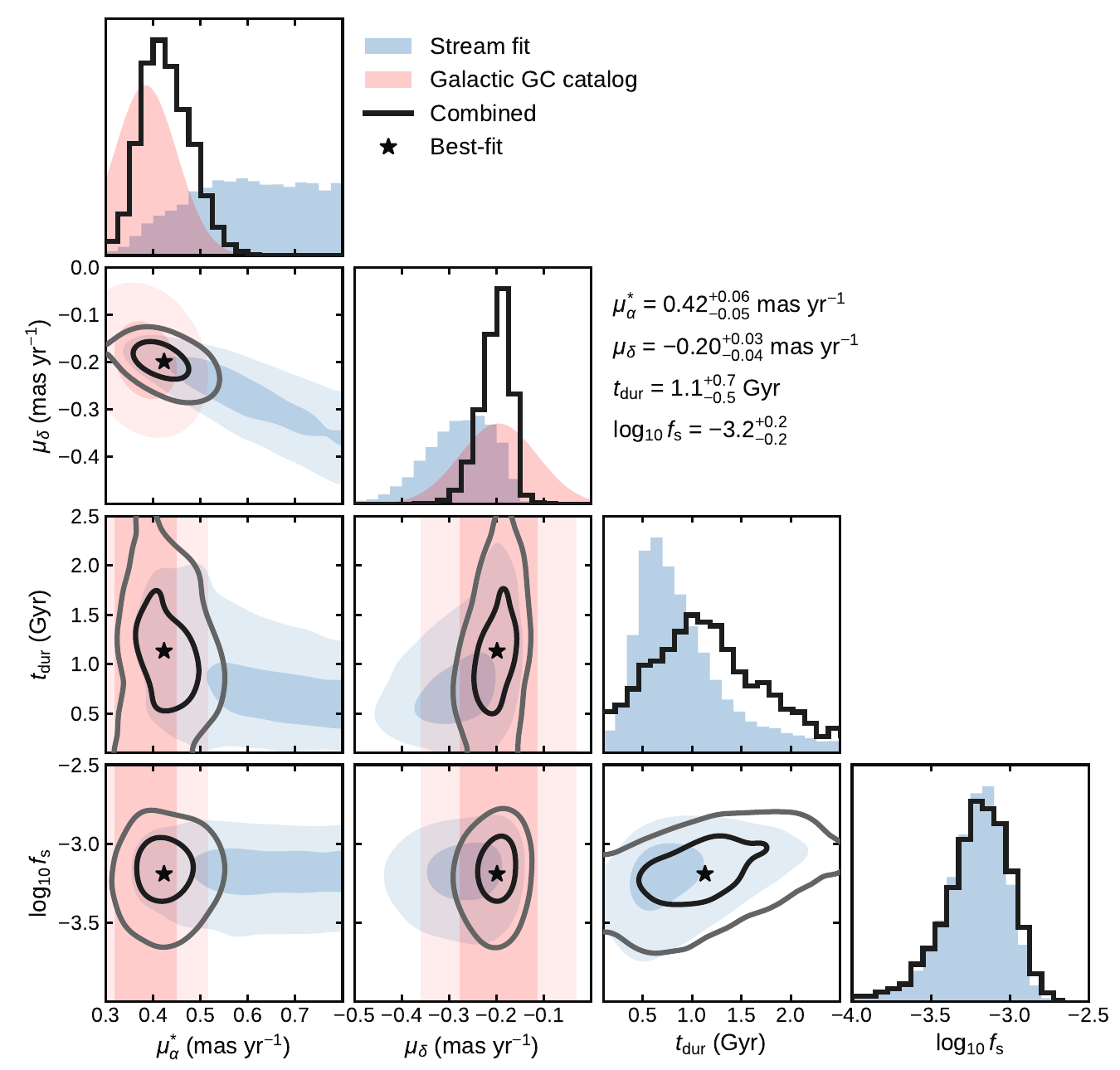}
    \vspace{-5mm}
    \caption{Posterior distribution of the stream parameters: the progenitor cluster's PM ($\mu_\alpha^*$, $\mu_\delta$), the stream release duration $t_{\rm dur}$, and the stream fraction $f_{\rm s}$. Blue shows the posterior from the stream fit alone, pink shows the PM from the Galactic GC database, and black shows the combined posterior, with the median values marked as star symbols.}
    \label{fig:corner}
\end{figure*}}
\iflatexml\else\FigCorner\fi

\section{Stream detection method}
\label{sec:method}

We use the framework of \textit{StarStream} \citep{chen_starstream_2025} to detect the stellar stream around AM-1. \textit{StarStream} models stars around a GC as a mixture of stream and background. The stream component is modeled by a particle-spray algorithm \citep{chen_improved_2025}, while the observed data determine the background component. Both components include the uncertainties of each star and are normalized within the search region. Following \citet{chen_starstream_2025}, the log-likelihood of the $N$ stars is
\begin{equation}
    \ln \mathcal{L} \equiv \sum_{i=1}^{N} \ln \left[ f_{\rm s}\, p_{{\rm s},i} + (1 - f_{\rm s})\, p_{{\rm bg},i} \right],
    \label{eq:likelihood}
\end{equation}
where $f_{\rm s}$ is the stream fraction, and $p_{{\rm s},i}$ and $p_{{\rm bg},i}$ are the stream and background densities of the $i$-th star. We measure the evidence for a stream by $\Delta \ln \mathcal{L}$, the log-likelihood gain relative to the background-only ($f_{\rm s} = 0$) model. We compute the membership probability of each star as in \citet{chen_starstream_2025}:
\begin{equation}
    P_{{\rm s},i} = \frac{f_{\rm s}\, p_{{\rm s},i}}{f_{\rm s}\, p_{{\rm s},i} + (1 - f_{\rm s})\, p_{{\rm bg},i}}.
    \label{eq:membership}
\end{equation}

Since \textit{StarStream} was originally designed for \textit{Gaia}, we adapt it to \textit{Euclid} in \S\ref{sec:method_adapt}. To account for the uncertain PM and unknown stream release duration $t_{\rm dur}$, we also treat them as free parameters like $f_{\rm s}$. In \S\ref{sec:method_mcmc}, we describe the MCMC approach we employ to fit these parameters.

\newcommand{\FigDetection}{%
\begin{figure*}
    \centering
    \includegraphics[width=\linewidth]{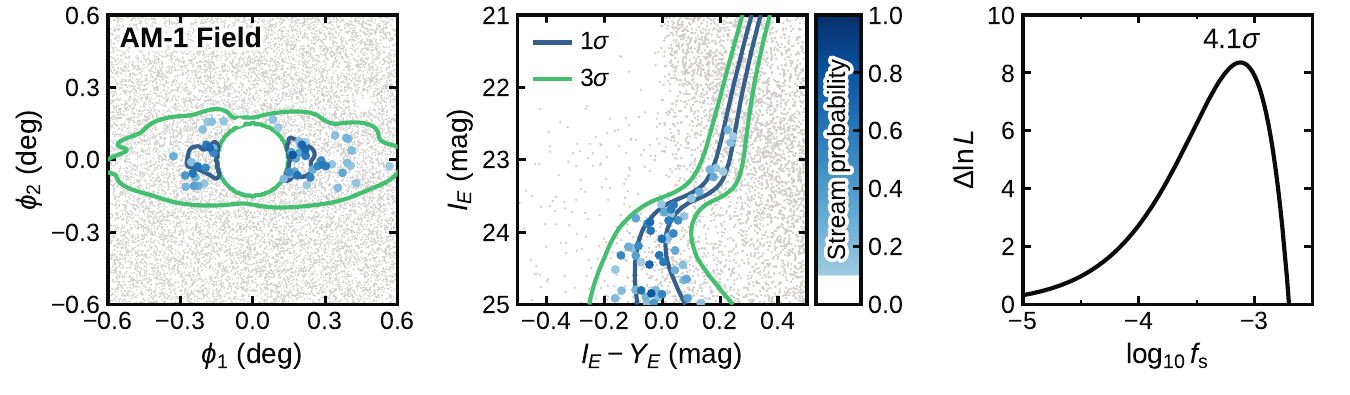}\\
    \includegraphics[width=\linewidth]{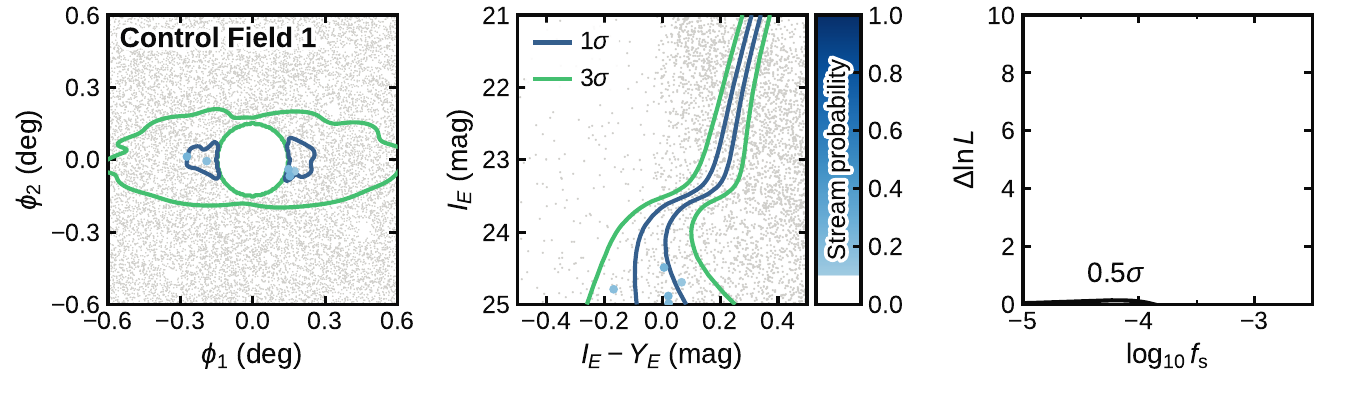}
    \vspace{-5mm}
    \caption{Stream detection in the AM-1 field (\textit{top row}) and Control Field 1 (\textit{bottom row}). \textit{Left:} sky positions in the $\phi_1$--$\phi_2$ coordinates. The white void is the masked tidal radius. \textit{Middle:} CMD. In both panels, gray points represent all stars in the search region. Blue points are detected stream members, color-coded by their membership probability $P_{\rm s}$. To guide the eye, we also show the $1\sigma$ and $3\sigma$ contours of the stream model in the \textit{left panel}, while the curves in the \textit{middle panel} represent the $1\sigma$ and $3\sigma$ boundaries around the isochrone. \textit{Right:} log-likelihood gain $\Delta \ln \mathcal{L}$ as a function of the stream fraction $f_{\rm s}$, labeled with the formal significance at its peak.
    }
    \label{fig:am1_stream_detection}
\end{figure*}}
\iflatexml\else\FigDetection\fi

\subsection{Adaptation to \textit{Euclid}}
\label{sec:method_adapt}

We make two main changes. First, DR1 has no PMs, and most stream stars are too faint for \textit{Gaia}. We therefore only search the 4D space of sky positions, color ($I_E - Y_E$), and magnitude ($I_E$). Second, we replace the original Gaussian kernel density estimation (KDE) method with Gaussian mixture models (GMM) to construct both the stream and background models. In principle, Gaussian KDE is a special case of GMM. Therefore, GMM preserves the advantages of Gaussian KDE: the convolution with the measurement uncertainties is analytic, because the covariances of each component and each star add, and the density normalizes exactly within a truncated search region. However, a GMM needs only several tens of components for the whole field instead of one per each star. This makes it much more efficient than KDE, which is important for MCMC. We have verified that the two methods have similar outputs as long as the GMM has a sufficient number of components.

We use the $\phi_1$--$\phi_2$ coordinate system, which is centered on AM-1 and aligned with the PM from the Galactic GC database (Table~\ref{tab:am1}). Note that a GMM with full covariance does not depend on the orientation of the coordinate system, so this alignment is mainly for visual clarity and does not affect the final results. We then model the stream density as the product of a sky distribution and a distribution in the color--magnitude diagram (CMD). This assumes that all stream stars share the same distance. This assumption is safe, since our particle-spray models predict a distance modulus scatter of only 0.02~mag in the search region, smaller than the photometric uncertainties.

For the sky distribution, we run the particle-spray algorithm of \citet{chen_improved_2025} with \texttt{Agama} \citep{vasiliev_agama_2019} in the \texttt{MilkyWayPotential2022} model \citep{price-whelan_gala_2017}. We release a fixed number of 12,000 particles at a constant rate over a time interval $t_{\rm dur}$. The fixed number of particles ensures that the noise of the stream model does not depend on $t_{\rm dur}$. We then fit the particles in the search region with a 2D GMM of 32 components. We have verified that varying the number of components between 16 and 64 does not significantly alter the results. Too few components underfit the true distribution of particles, while too many overfit it. Both reduce the detection accuracy.

For the CMD distribution, we use a MESA Isochrones and Stellar Tracks (MIST) isochrone \citep{choi_mesa_2016, dotter_mesa_2016} with the age, $[{\rm Fe/H}]$, and $A_V$ in Table~\ref{tab:am1}, placed at the cluster distance. We shift the model colors by $-0.018$~mag to better fit AM-1's red giants. This shift is needed because the \citet{weisz_dolphoteuclid_2026} parameters come from fits of other stellar models to archival \textit{HST} photometry. \citet{weisz_dolphoteuclid_2026} also found that these models are offset from the \textit{Euclid} photometry, which they attributed to errors in the $Y_E$ zero point or in the pre-launch NISP passbands. We represent the isochrone as a GMM in the CMD. Each component is a Gaussian centered on one point along the isochrone. We space these points by 0.01~mag and weight each component by the \citet{kroupa_variation_2001} initial mass function. The isochrone itself is a curve with no width. We therefore give each component a small intrinsic width (0.002~mag in color and 0.01~mag in magnitude), which is much smaller than the photometric uncertainties.

For the background, we use a GMM of 64 components in the full 4D space. We fit it to all stars in the search region using extreme deconvolution \citep{bovy_extreme_2011}, which is detailed in Appendix~\ref{sec:gmm}. The fit gives the intrinsic distribution of background stars, which we then convolve with the uncertainties of each star when evaluating the likelihood. A joint 4D model allows the background CMD to vary across the search region. A held-out test does not clearly prefer any number of components between 16 and 64.

When evaluating the likelihood, we convolve each GMM component with the photometric uncertainties of each star and normalize the result within the search region and $I_E<25$. We take the uncertainties from the MER catalog and add a floor of 0.01~mag in quadrature to each band to account for unmodeled systematic uncertainties. We treat sky coordinates as exact, since the astrometric uncertainties are negligible on the angular scales of the stream.

\subsection{MCMC sampling}
\label{sec:method_mcmc}

We sample the cluster PM, $t_{\rm dur}$, and $f_{\rm s}$ with \texttt{emcee} \citep{ForemanMackey2013}. We employ uniform priors in $\mu_\alpha^* \in [0.3, 0.8]\ {\rm mas\,yr^{-1}}$, $\mu_\delta \in [-0.5, 0]\ {\rm mas\,yr^{-1}}$, $t_{\rm dur} \in [0.1, 3]$~Gyr, and $\log_{10}f_{\rm s} \in [-5, -1]$. Each step requires a new particle-spray run of the stream, which takes 5--10~s. The CMD distribution of the stream and the background model do not depend on the free parameters, so we compute them only once. We run 48 walkers for 2,000 iterations and discard the first 65 iterations (three times the longest autocorrelation time) as burn-in. The chains are around 90 autocorrelation times long, well above the $\approx10$ recommended by \citet{ForemanMackey2013}.

The stream alone can constrain the PM. This constraint is independent of the PM from the Galactic GC database. We therefore also combine the two by importance resampling: we weight each posterior sample by the likelihood of its PM under the database measurement. The prior range covers $\sim90\%$ of the probability of the database PM, so the combined posterior is slightly truncated. We report both posteriors in \S\ref{sec:results}.

\section{Results}
\label{sec:results}

We show the posterior distribution of stream parameters in Fig.~\ref{fig:corner} (blue contours). The stream fraction $f_{\rm s}\approx10^{-3.2}$ does not correlate with other parameters, indicating that the detected signal is insensitive to the exact stream parameters. On the other hand, the linear correlation between $\mu_\alpha^*$ and $\mu_\delta$ is expected since the PM direction sets the orientation of the stream on the sky. A posterior distribution concentrated along this direction supports the existence of a stream-like, directional overdensity, rather than an isotropic envelope. There is a degeneracy between the PM and the release duration $t_{\rm dur}$. This is because a faster-moving progenitor needs a shorter $t_{\rm dur}$ to create the same stream length.

\iflatexml\FigCorner\fi

We then combine the MCMC results with the PM from the Galactic GC database (pink contours in Fig.~\ref{fig:corner}). The combined posterior prefers $\mu_\alpha^*\approx0.42\ {\rm mas\,yr^{-1}}$ and $\mu_\delta\approx-0.2\ {\rm mas\,yr^{-1}}$. In fact, the database PM also lies along the same direction preferred by the stream fit. Therefore, our stream fit does not conflict with prior measurements. Instead, it adds new information by constraining the cluster's motion using an independent measurement.

Since the combined PM has narrower ranges, this also helps constrain $t_{\rm dur}$ to around 1.1~Gyr. Such a duration is about half of the radial orbital period of AM-1, which was at its pericenter around that time. Therefore, if AM-1 indeed was accreted, it must have been accreted more than $t_{\rm dur}$ ago and have experienced at least one pericenter passage. In Table~\ref{tab:am1}, we also list the values and uncertainties from the combined fit.

Based on these parameters, we show the detection result in the \textit{top row} of Fig.~\ref{fig:am1_stream_detection}. The log-likelihood gain peaks at $\Delta \ln \mathcal{L} \approx 8.4$, which corresponds to a formal significance of $\sqrt{2\,\Delta \ln \mathcal{L}} = 4.1\sigma$ for one free parameter \citep[$f_{\rm s}$;][]{Wilks1938}. This detection reveals 15 stars with stream membership probability $P_{\rm s}>0.5$, and 55 stars with $P_{\rm s}>0.1$.

\iflatexml\FigDetection\fi

On the sky, we find two tidal tails extending outside the tidal radius along the $\phi_1$ direction. Each tail only extends around $0.3$~deg from the cluster center, but with a width $\approx0.1$~deg. Such short, puffy tails may not look like a stream, but they are reproduced by our particle-spray model. The model predicts this shape because AM-1 is near its apocenter, where the stream stars are compressed along the orbit. In the CMD, most detected stars lie near or below the MSTO, where the stream-to-background ratio is the highest. This behavior was also found by \citet{chen_starstream_2025}.

To test whether the extra-tidal stars form tidal tails rather than an isotropic envelope, we replace the sky density of the stream model with a radially symmetric Plummer profile, and sample its scale radius with MCMC. This reduces $\Delta \ln \mathcal{L}$ from 8.4 to 4, so the stream model is preferred.

\begin{figure}
    \centering
    \includegraphics[width=\linewidth]{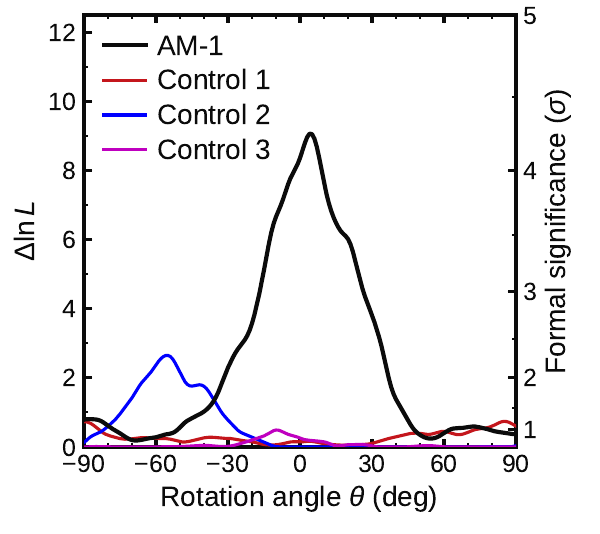}
    \vspace{-5mm}
    \caption{Null tests. Log-likelihood gain $\Delta \ln \mathcal{L}$ as a function of the field rotation angle $\theta$, for the AM-1 field (black) and Control Fields 1--3 (red, blue, and magenta). We also mark the corresponding formal significance on the right axis. $\theta=0$ corresponds to the original data.
    }
    \label{fig:rotation_test}
\end{figure}

To confirm this is a real detection instead of a false positive, we apply the same stream model to the three control fields as null tests. This is done by performing rotation tests in all four fields. For each field, we rotate all stars about the field center by an angle $\theta$ from $-90$ to $90$~deg, while keeping the stream model fixed. We then recompute the maximum $\Delta \ln \mathcal{L}$ at each rotation angle. As shown in Fig.~\ref{fig:rotation_test}, $\Delta \ln \mathcal{L}$ of the AM-1 field peaks near $\theta=0$ and drops below unity beyond $\theta\approx\pm45$~deg. This again shows that the signal comes from a directional overdensity aligned with the PM. In contrast, the control fields show weaker signals. Control Fields 1 and 3 stay below $\Delta \ln \mathcal{L}=1$ at all angles, and Control Field 2 reaches at most $\Delta \ln \mathcal{L}\approx2.6$ (or $2.3\sigma$) at $\theta\approx-55$~deg, likely due to a background fluctuation that happens to align with the stream model.

As an example, we show Control Field 1 without rotation in the \textit{bottom row} of Fig.~\ref{fig:am1_stream_detection}. Its $\Delta \ln \mathcal{L}$ peaks at only 0.1 (or $0.5\sigma$), at a lower $f_{\rm s}$, and no star has $P_{\rm s}>0.5$. Across all rotation angles, the three control fields have at most 3 stars with $P_{\rm s}>0.5$. According to the 10-star threshold in our previous work \citep{chen_starstream_2025,chen_improved_2025}, all three would be considered non-detections. In contrast, the AM-1 field passes this threshold with 15 stars. We therefore conclude that we have found AM-1's stellar stream in \textit{Euclid} DR1.

Our null tests support that the signal in the AM-1 field is real rather than random noise. Strictly speaking, however, a real signal may still come from a background fluctuation that happens to resemble a stream. The rotation tests partly address this scenario: across all rotation angles, the significance reaches at most $2.3\sigma$. A strict test would require sampling all stream parameters with MCMC in each control field, which is beyond the scope of this work.

\section{Discussion}
\label{sec:discussion}

\subsection{A matched-filter view of the stream}
\label{sec:window}

To check that the detection does not rely on the \textit{StarStream} framework, we also apply a simple matched-filter method \citep[e.g.,][]{rockosi_matched-filter_2002}. We select stars with $23 < I_E < 25$ and within $2\sigma$ of the isochrone in the CMD, where $\sigma$ is the observed color uncertainty. We obtain $\sigma$ as a function of $I_E$ by taking the median color uncertainty of all stars in 0.25-mag bins and interpolating linearly. We then estimate the sky density of the selected stars with a Gaussian KDE with a bandwidth of 0.15~deg.

Fig.~\ref{fig:band_kde_am1_vs_null} shows the density maps of the AM-1 field and Control Field 1, both normalized by the mean density in Control Field 1. The selected stars show an overdensity of up to a factor of $\approx2$--3 around the cluster. This overdensity is elongated along $\phi_1$, consistent with the tidal tails in Fig.~\ref{fig:am1_stream_detection}. In Control Field 1, the density fluctuations are weaker and show no coherent structure around the center. Therefore, the overdensity around AM-1 is visible even without \textit{StarStream}.

\newcommand{\FigMatched}{%
\begin{figure}
    \centering
    \includegraphics[width=\linewidth]{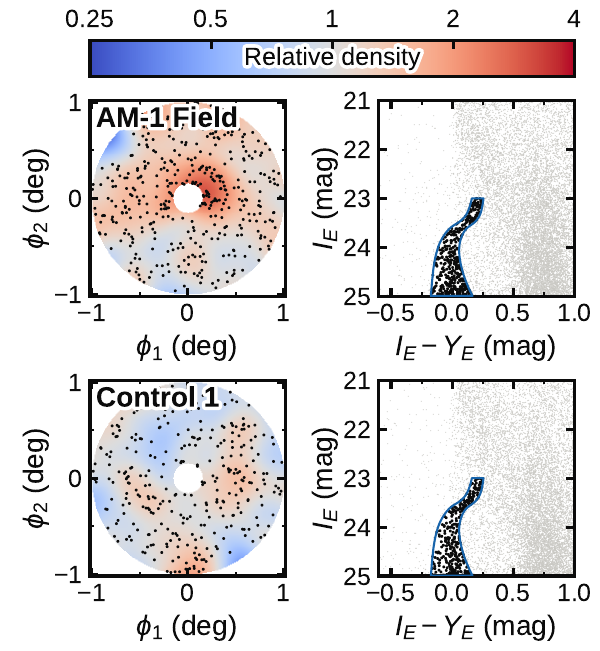}
    \vspace{-5mm}
    \caption{Matched-filter view of the AM-1 field (\textit{top row}) and Control Field 1 (\textit{bottom row}). \textit{Left:} sky positions of stars within the selection window. The color scale represents the number density of stars via KDE (relative to the mean density in Control Field 1). \textit{Right:} Selection window in the CMD: $23<I_E<25$ and within $2\sigma$ of the isochrone (blue contour). Selected stars are shown in black. 
    }
    \label{fig:band_kde_am1_vs_null}
\end{figure}}
\iflatexml\FigMatched\fi

However, a simple matched filter cannot quantify how significant this overdensity is. We need either a matched filter with a significance estimate \citep[e.g.,][]{bernard_serendipitous_2014,shipp_stellar_2018} or \textit{StarStream}. \textit{StarStream} turns this visual impression into a single detection significance value, and it also assigns a membership probability to each star. Such automated, quantitative methods are thus necessary for stream detection in upcoming wide-field surveys, such as \textit{Euclid} itself, the \textit{Rubin} Legacy Survey of Space and Time \citep[LSST;][]{ivezic_lsst_2019}, and the \textit{Roman} High Latitude Wide Area Survey \citep[HLWAS;][]{spergel_wide-field_2015}, where visual inspection becomes impractical.

\iflatexml\else\FigMatched\fi

\subsection{Mass loss rate of AM-1}
\label{sec:mdot}

Stellar streams consist of stars that their progenitors have lost, so they directly probe the dynamical evolution of the progenitors. We convert the detected stream stars into the mass loss rate of AM-1 with equation~(4) of \citet{chen_starstream_2026}:
\begin{equation}
    \dot{M} \approx f_{\rm detect}^{-1}\, \frac{\dot{N}_{\rm tracer}\, M_{\rm obs}\, w_{\rm GC}}{N_{\rm tracer}}.
    \label{eq:mdot}
\end{equation}
Here, $\dot{N}_{\rm tracer}$ is the rate at which the particle-spray model releases particles, and $N_{\rm tracer}$ is the number of these particles in the search region. $M_{\rm obs} = 12.4\,M_\odot$ is the summed mass of the 15 members with $P_{\rm s} > 0.5$, where we take the mass of each star from the isochrone. The factor $w_{\rm GC} = 9.7^{+3.1}_{-2.1}$ accounts for stream stars below our detection limit ($0.78\,M_\odot$ at $I_E = 25$). We obtain it by extrapolating the present-day mass function slope of AM-1 ($\alpha = -1.07 \pm 0.34$; see Table~\ref{tab:am1}) down to $0.08\,M_\odot$. Finally, $f_{\rm detect} \equiv N_{\rm detect}/N_{\rm true}$ is the ratio between stars detected by \textit{StarStream} with $P_{\rm s} > 0.5$ and the true number of stream stars.

\citet{chen_starstream_2026} found $f_{\rm detect} = 0.9$ for \textit{Gaia} using the mock stream catalog by \citet{holm-hansen_catalog_2026}. This work, however, uses four observables instead of six and relies on faint stars, so we recalibrate $f_{\rm detect}$ here with mock streams of AM-1 (detailed in Appendix~\ref{sec:recovery}). For $N_{\rm detect} = 15$, our calibration yields $N_{\rm true} = 20^{+6}_{-5}$ (Fig.~\ref{fig:member_recovery_poisson}), i.e., $f_{\rm detect} \approx 0.75$. We then propagate the Poisson noise of the stream stars, the uncertainties of $f_{\rm detect}$, the mass function slope $\alpha$, and the cluster age to calculate the uncertainty of $\dot{M}$. We obtain $\dot{M} = 0.24^{+0.12}_{-0.09}\ M_\odot\,{\rm Myr^{-1}}$, averaged over $t_{\rm dur}\approx1.1$~Gyr. 

At first glance, this value seems low compared with the typical $\dot{M}\sim1\ M_\odot\,{\rm Myr^{-1}}$ for Galactic GCs \citep{kupper_globular_2015,gieles_supra-massive_2021,chen_stellar_2025,chen_starstream_2026}. However, our value is consistent with the present-day mass of AM-1 divided by its dissolution time in the Galactic GC database, $0.15\,M_\odot$\,Myr$^{-1}$. Note that the database estimates this time using recipes calibrated on $N$-body simulations \citep{baumgardt_dynamical_2003}. Therefore, finding a stream around such a distant GC may be surprising, but the existence of the stream itself is not. The stream of AM-1 was not found before simply because previous data were not deep enough. Of the $N_{\rm detect}=15$ members, none is in \textit{Gaia} and only four are brighter than the $10\sigma$ depth of DES ($i\approx23.8$).

Three effects may bias our estimate. First, we may overestimate $f_{\rm detect}$ since a real stream can deviate from the model. Second, mass segregation preferentially removes low-mass stars, so the stream may contain more low-mass stars than the extrapolated mass function predicts. Third, the point-source selection may remove $\sim10\%$ of real stars, leading to an underestimated $\dot{M}$.

\newcommand{\FigGMM}{%
\begin{figure*}
    \centering
    \includegraphics[width=\linewidth]{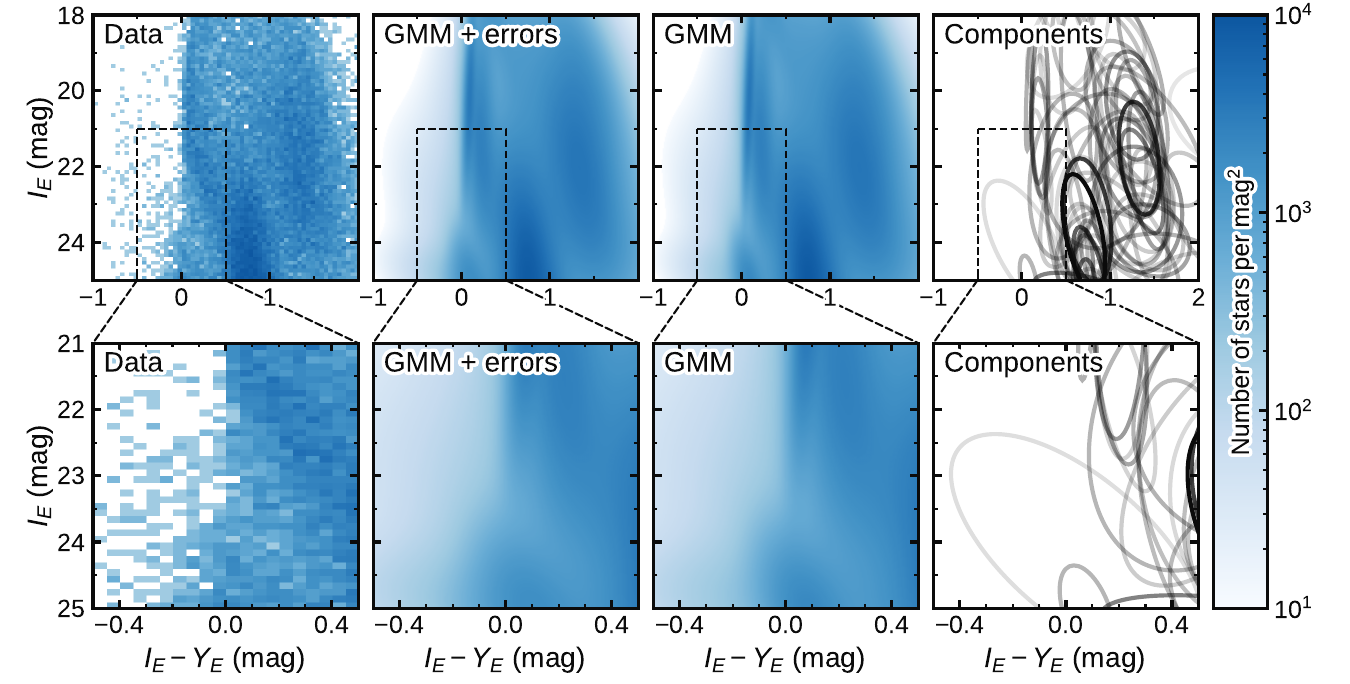}
    \vspace{-5mm}
    \caption{Background GMM in the CMD. \textit{Top row:} Full CMD of the search region. \textit{Bottom row:} Zoom-in regions near the AM-1 isochrone. From \textit{left} to \textit{right}: true CMD distribution of catalog sources (``Data''), reconstructed distribution (``GMM + errors''), intrinsic distribution (``GMM''), and individual Gaussian components (``Components'', shown as $1\sigma$ ellipses). The color scale shows the number density of stars.}
    \label{fig:am1_background_gmm}
\end{figure*}}
\iflatexml\else\FigGMM\fi

\section{Summary}
\label{sec:summary}

In this work, we search for a stellar stream around AM-1 in \textit{Euclid} DR1 imaging of EDF-South (\S\ref{sec:data}). AM-1 is one of the most distant Galactic GCs at $D_\odot\approx 120$~kpc. We adapt the \textit{StarStream} method to \textit{Euclid} in the 4D space of sky position, color ($I_E - Y_E$), and magnitude ($I_E$). This method constructs both the stream and background models with GMMs, which are then convolved with the photometric uncertainties (\S\ref{sec:method}). 

We detect a stream around AM-1 at $\Delta\ln L=8.4$, or a formal significance of $4.1\sigma$ (\S\ref{sec:results}). We find 15 member stars with $P_{\rm s} > 0.5$. The stream has two short ($\approx 0.3$~deg) and wide ($\approx 0.1$~deg) tails along the PM direction, and most members lie near or below the MSTO (Fig.~\ref{fig:am1_stream_detection}). An isotropic Plummer profile in place of the stream reduces $\Delta \ln \mathcal{L}$ from 8.4 to 4. In contrast, applying the same model to three control fields yields at most $2.3\sigma$ and 3 stars with $P_{\rm s} > 0.5$, below the 10-star threshold for a valid detection. A simple matched filter also reveals an elongated overdensity along the same direction (\S\ref{sec:window}, Fig.~\ref{fig:band_kde_am1_vs_null}).

The computational efficiency of the method enables fast sampling of the cluster PM and the stream release duration with MCMC (\S\ref{sec:method_mcmc}). This allows us to obtain new constraints on AM-1's PM: $\mu_\alpha^* \approx 0.42$ and $\mu_\delta \approx -0.20\ {\rm mas\,yr^{-1}}$, as well as a release duration $ t_{\rm dur}\approx 1.1$\,Gyr, about half of the radial orbital period (Fig.~\ref{fig:corner}, Table~\ref{tab:am1}).

We also infer a mass loss rate of $\dot{M} \approx 0.24\ M_\odot\,{\rm Myr^{-1}}$ (\S\ref{sec:mdot}). This rate appears low compared with typical Galactic GCs, but is consistent with theoretical predictions from $N$-body simulations.

These results show that the stream of AM-1 was missing not because the cluster is not losing mass, but because the stream stars lie near the MSTO at $I_E \approx 24$, beyond the reach of any other surveys in the vicinity of AM-1. In addition to \textit{Euclid} itself, \textit{Rubin} and \textit{Roman} will reach similar depths over large areas in multiple bands. Together with automated methods such as \textit{StarStream}, they will enable a systematic search for the streams of distant GCs throughout the MW halo.

\section*{Acknowledgments}
Y.C. acknowledges support from the CCAPP fellowship at The Ohio State University. 
AI tools (Claude) were used for code development and language improvement. We did not use AI tools for interpreting scientific conclusions.

\software{
\textit{StarStream} \citep{chen_starstream_2025},
\texttt{Agama} \citep{vasiliev_agama_2019},
\texttt{Gala} \citep{price-whelan_gala_2017},
\texttt{JAX} \citep{bradbury_jax_2018},
\texttt{emcee} \citep{ForemanMackey2013},
\texttt{Astropy} \citep{the_astropy_collaboration_astropy_2013,the_astropy_collaboration_astropy_2018,the_astropy_collaboration_astropy_2022},
\texttt{NumPy} \citep{harris_array_2020},
\texttt{SciPy} \citep{virtanen_scipy_2020},
\texttt{Matplotlib} \citep{hunter_matplotlib_2007},
\texttt{h5py} \citep{collette_python_2013},
\texttt{pandas} \citep{mckinney_data_2010,the_pandas_development_team_pandas-devpandas_2024}
}

\facility{
\textit{Euclid}
}

\newpage
\appendix
\vspace{-6mm}

\section{GMM}
\label{sec:gmm}

We use three GMMs in \S\ref{sec:method}: the background model, the sky distribution of the stream, and the CMD distribution of the stream. This appendix describes how we normalize, fit, and evaluate them.

A GMM with weights $w_k$, means $\boldsymbol{m}_k$, and covariances $\mathbf{V}_k$ describes the intrinsic distribution of the stars. For star $i$ with photometric covariance $\mathbf{S}_i$, the density within the search region $W$ is
\begin{equation}
    p_i = \frac{1}{P_i(W)} \sum_{k=1}^{K} w_k\, \mathcal{N}(\boldsymbol{m}_k, \mathbf{V}_k + \mathbf{S}_i),
    \label{eq:app_density}
\end{equation}
where $\mathcal{N}$ is evaluated at the observed values of star $i$, and $P_i(W)$ is the probability that the uncertainty-convolved GMM falls within $W$. Since the sky coordinates are exact, we convolve the sky and CMD distributions of the stream separately.

We fit each GMM by maximizing the likelihood of Equation~(\ref{eq:app_density}) for its training data. Without $P_i(W)$, this becomes the extreme deconvolution of \citet{bovy_extreme_2011}, while including $P_i(W)$ extends the GMM beyond $W$, and only the overlapping region is constrained by the data. Each covariance is a free Cholesky term plus a fixed minimum width along each axis. For the background, the minimum widths are 0.01~deg in sky coordinates, 0.002~mag in color, and 0.01~mag in magnitude. For the sky distribution of the stream, the minimum width is also 0.01~deg. These values match the widths of the stream model, so neither the stream nor the background can describe finer structure than the other. The minimum widths prevent components from collapsing onto single stars and set the finest scale the GMM can resolve. We initialize the fit with $k$-means++ \citep{arthur_k-means_2007} and 50 iterations of the extreme deconvolution algorithm.

As an example, Fig.~\ref{fig:am1_background_gmm} shows the background GMM in the CMD. In the ``GMM + errors'' panel, we convolve it with the photometric uncertainties, which are derived from the catalog uncertainties at the same $I_E$ and $Y_E$ in 0.25-mag bins.

\iflatexml\FigGMM\fi

\newcommand{\FigRecovery}{%
\begin{figure}
    \centering
    \includegraphics[width=\linewidth]{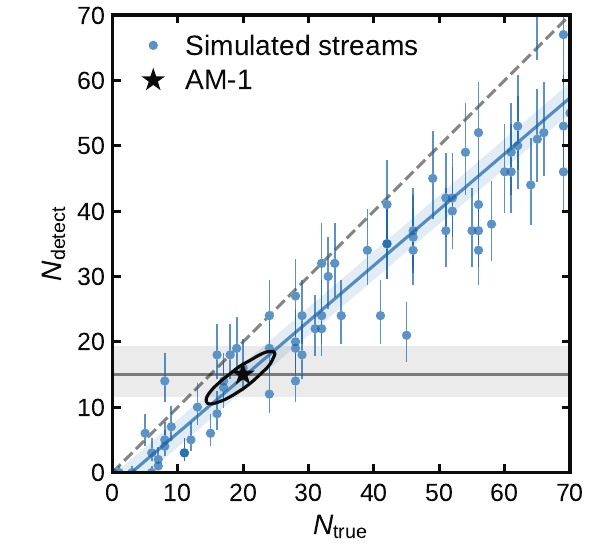}
    \vspace{-5mm}
    \caption{Recovery of stream stars in the 90 injection tests. Blue points show $N_{\rm detect}$ against $N_{\rm true}$, with the 16th--84th percentile Poisson range of each count. The blue line and band show the fitted relation and its intrinsic scatter; the dashed line represents $N_{\rm detect} = N_{\rm true}$; and the gray line and band mark AM-1's $N_{\rm detect}=15$. We show the inferred $N_{\rm true}$ of AM-1 as the black contour ($1\sigma$ credible region) and star (median).
    }
    \label{fig:member_recovery_poisson}
\end{figure}}
\iflatexml\else\FigRecovery\fi

\section{Member recovery efficiency}
\label{sec:recovery}

We calibrate the recovery efficiency of stream stars with mock streams of AM-1 injected into the three control fields. Each mock stream is a new particle-spray realization using the best-fit stream parameters in \S\ref{sec:results}. We populate it with stars drawn from the mass function of AM-1 (whose power-law slope is from the Galactic GC database). Each mock star takes the flux uncertainties of a random star in this control field with similar $I_E$ (within 0.1~mag), and we draw its observed fluxes from Gaussians with these uncertainties. We create 30 mock streams for each control field with mass loss rates uniformly spaced in 0--1~$M_\odot\,{\rm Myr}^{-1}$. This totals 90 mock streams. For each stream, we rotate the background by a random angle.

We then analyze each mock in the same way as the AM-1 field. In Fig.~\ref{fig:member_recovery_poisson}, we compare two numbers for each test. $N_{\rm true}$ is the true number of injected stars in the search region with $I_E < 25$, and $N_{\rm detect}$ is the number of stars revealed by \textit{StarStream} with $P_{\rm s} > 0.5$.

\iflatexml\FigRecovery\fi

Besides Poisson noise, $N_{\rm detect}$ has additional scatter from the background and from the detection method itself, which we denote $\sigma_{\rm int}$. To account for both, we model $N_{\rm detect}$ as a Poisson variable whose mean follows $\mu \sim \mathcal{N}(a + bN_{\rm true}, \sigma_{\rm int}^2)$. We then sample $(a, b, \sigma_{\rm int})$ with \texttt{emcee} and flat priors. The fit gives $b = 0.86^{+0.02}_{-0.02}$, $a = -2.5^{+0.7}_{-0.8}$, and $\sigma_{\rm int} = 2.1^{+0.7}_{-0.6}$. 

The simplest way to infer AM-1's $N_{\rm true}$ from its $N_{\rm detect}=15$ is to invert the fitted relation. However, this ignores two sources of uncertainty. First, the same $N_{\rm true}$ can give different $N_{\rm detect}$ because of Poisson noise and $\sigma_{\rm int}$. Second, $a$, $b$, and $\sigma_{\rm int}$ are themselves uncertain. We therefore infer the posterior distribution of $N_{\rm true}$ given $N_{\rm detect} = 15$. For each $N_{\rm true}$, we compute the probability of detecting 15 stars under the fitted relation, and we average it over the posterior samples of $(a, b, \sigma_{\rm int})$. We adopt a log-uniform prior on $N_{\rm true}$ over the range of the tests. This gives $N_{\rm true} = 20^{+6}_{-5}$ (black contour and star in Fig.~\ref{fig:member_recovery_poisson}).

\bibliographystyle{aasjournal}
\bibliography{library}

\end{document}